# *Effect of TeO$_2$ addition on the dielectric properties of CaCu$_3$Ti$_4$O$_{12}$ ceramics derived from the oxalate precursor route.*

P.Thomas[a], and K.B.R.Varma.[b*]

[a] *Dielectric Materials Division, Central Power Research Institute, Bangalore:560080, India*

[b] *Materials Research Centre, Indian Institute of Science, Bangalore: 560012, India*

Abstract:

CaCu$_3$Ti$_4$O$_{12}$ (CCTO) ceramics which has perovskite structure gained considerable attention due to its giant permittivity. But, it has high tan δ (0.1 at 1kHz) at room temperature, which needs to be minimised to the level of practical applications. Hence, TeO$_2$ which is a good glass former has been deliberately added to CCTO nano ceramic (derived from the oxalate precursor route) to explore the possibility of reducing the dielectric loss while maintaining the high permittivity. The structural, morphological and dielectric properties of the pure CCTO and TeO$_2$ added ceramics were studied using X-ray diffraction, Scanning Electron Microscope along with Energy Dispersive X-ray Analysis (EDX), spectroscopy and Impedance analyzer. For the 2.0 wt % TeO$_2$ added ceramics, there is a remarkable difference in the microstructural features as compared to that of pure CCTO ceramics. This sample exhibited permittivity values as high as 7387 @ 10 KHz and low dielectric loss value of 0.037 @ 10 kHz, which can be exploited for the high frequency capacitors application.

**1.0 Introduction**

Materials with high permittivity are in great demand for the miniaturization of electronic devices. The titanate compound CaCu$_3$Ti$_4$O$_{12}$ (CCTO) belongs to a family of the type, ACu$_3$Ti$_4$O$_{12}$ (where A= Ca or Cd) [1], which has centrosymmetric *bcc* structure (space group *Im$^3$*, lattice parameter *a=7.392A$^o$*) has gained considerable attention due to its large permittivity (ε~10$^{4-5}$) which is nearly independent of frequency (upto 10 MHz) and temperature (upto 300°C) [2,3]. In order to understand the origin of giant dielectric response in CCTO, extensive work has been carried out and various mechanisms have been proposed [2-26]. But, the generally accepted mechanism is an internal barrier layer capacitor (IBLC) model, which consists of semi conducting grains and insulating grain boundaries. However, the dielectric loss (tan $\delta$) of CCTO is high for the practical applications. The typical value of tan $\delta$ is around 0.1 at 1kHz at room temperature. It is very essential to reduce the dielectric loss in CCTO and hence, various attempts have been made to decrease the dielectric loss [27-43]. The effective way to decrease the dielectric loss in CCTO is to increase the resistance of the grain boundary by cation doping, thereby one can control the chemistry and structure of interfacial regions at grain boundaries [27-43]. Extensive work has been carried out on the effect of doping on the dielectric properties in CCTO [27-45], and it has been observed that the doping agents had reduced the dielectric loss to a greater extent while maintaining its high permittivity [30-43]. The calcium stoichiometry in CCTO exhibited high permittivity accompanied by low loss due to the low level of CuO segregation at the grain boundaries [46]. Hence, controlling the CuO segregation at the grain boundaries would give rise to improved dielectric properties. This could be achieved by employing glasses that has high electrical resistance and low loss [47,48], that could assist in the lowering of loss and in improving the overall electrical properties of the polycrystalline ceramics. Considering this aspect in mind, $TeO_2$, which is a good network former and has low melting point (T=733°C) has been chosen, deliberately added to the CCTO ceramics and that it might improve the dielectric properties by controlling the chemistry and structure of interfacial regions especially at the grain boundaries in CCTO. The recent work on the $TeO_2$ addition upto 1.5 % by weight in CCTO prepared by the solid state reaction route, decreased the dielectric los as low as 0.09 and exhibited bimodal grain distribution [49]. However, it is well documented that the dielectric properties of CCTO ceramics are affected by the ceramic processing conditions including sintering temperatures as well as the ambience. Hence, CCTO ceramics derived from the oxalate route has been chosen to study the effect of $TeO_2$ addition on the dielectric properties of CCTO and its role in controlling/ modifying the microstructure which helps in reducing the loss to a level that is suitable for the practical applications.

---

[*] Corresponding author : Tel. +91-80-2293-2914; Fax: +91-80-2360-0683.
E-mail : kbrvarma@mrc.iisc.ernet.in (K.B.R.Varma)

This article reports the details pertaining to the effect of $TeO_2$ addition on the dielectric properties of CCTO ceramics fabricated employing the powders obtained by the oxalate precursor route.

## 2.0 Experimental

Ceramic samples of $CaCu_3Ti_4O_{12}$ were prepared from the oxalate precursor route [18]. Initially, the titania gel was prepared from the aqueous $TiOCl_2$(0.05M) by adding $NH_4OH$ (aq) (at $25^oC$) till the pH reaches ~ 8.0 and washed off $NH_4Cl$ on the filter funnel. To this titania gel [0.4 moles $TiO_2xH_2O$ (where 92<x <118)], powdered $H_2C_2O_4 \bullet 2H_2O$ was added and mixed thoroughly without the addition of water. To the clear solution obtained, calcium carbonate was added and stirred. The solution remained clear without any precipitate formation. This solution was cooled to $10^oC$, and cupric chloride dissolved in acetone+water (80/20) was added and stirred continuously. The thick precipitate was separated, washed several times with acetone to make it chloride-free and dried in air. The precipitate thus prepared were isothermally heated above $680^oC$ to get the ceramic powders of $CaCu_3Ti_4O_{12}$. The phase pure CCTO nano powder was then mixed with different proportions of $TeO_2$. The $TeO_2$ content of 0.5,1.0,1.5 and 2.0 % by weight were added to CCTO powder. The resultant powder was ground thoroughly, ball milled for 2h and granulated by adding Polyvinylalcohol (PVA) and Poly ethylene glycol (PEG), then pressed into pellets (150MPa) with a diameter of 12mm. The green pressed pellets were slowly heated to $600^oC$ to get rid of the binder. Finally, the pellets were sintered in air at 1100 and $1130^oC$ for 2h. Pellet densities were measured by the Archimedes principle using xylene as the liquid medium. After sintering, X-ray powder diffraction studies were carried with an X'pert diffractometer (Philips, Netherlands) using Cu K$\alpha_1$ radiation ($\lambda$ = 0.154056 nm) in a wide range of 2θ (5° ≤ 2θ ≤ 85°) with 0.0170 step size to examine the phase constitutes of the specimens. Scanning electron microscope (FEI-Technai SEM-Sirion) equipped with Energy-Dispersive X-ray spectroscopy (EDX) capability was used to observe the microstructure and the composition of the sintered pellets. The capacitance measurements on the electroded (silver) pellets were carried out as a function of frequency (100Hz–1MHz) using an impedance gain-phase analyzer (HP4194A) at signal strength of 0.5Vrms. The measurement accuracy of the instrument is less than 5%. The dielectric permittivity was evaluated using the standard relation $\varepsilon_r = C \times d / \varepsilon_o A$, where $C$ =capacitance, $d$ is the thickness of the sample, $\varepsilon_o$ = 8.854X10$^{-12}$ F/m and A is the effective area of the sample.

### 3.0 Results and Discussion

The X-ray diffraction patterns were recorded on the as prepared CCTO powders (fig.1a) and on the pellets sintered at 1100°C (fig.1b). The X-ray patterns (Fig.1a&b) are compared well with ICDD data (01-075-1149) (fig.1(d)), demonstrating the single phase nature of CCTO. The TeO$_2$ added CCTO ceramics that were sintered at 1100°C/2h were also subjected to X-ray diffraction analysis (not shown here). The X-ray powder diffraction pattern obtained has revealed that there are secondary phases pertaining to TiO$_2$ and CaTiO$_3$, that could be detected. However, no secondary phases pertaining to either CuO or TeO$_2$ could be detected. Since there are secondary phases and also the pellets did not exhibit reasonable densities, keeping the dielectric properties in view, these samples were sintered at high temperatures (1130°C/2h). The X-ray powder diffraction patterns obtained for pure CCTO (fig.1c) and TeO$_2$ (0.5,1.0,1.5 and 2.0% TeO$_2$ by weight) added samples (fig.2(a-d)) that were sintered at 1130°C/2h are in consistent with that of the samples sintered at 1100°C/2h. The results obtained in this work is in contrast to what has been reported, wherein, upto 4% weight of TeO$_2$ addition, the XRD pattern showed single phase nature of CCTO [49]. The other doping studies [27-33] also reported the similar results, indicating that the addition to a certain levels exhibited single phase nature in CCTO. However, the work done on the CCTO modified by SrTiO$_3$ (ST) [33], it has been reported that the addition of SrTiO$_3$ exhibited mixed phases of CCTO and ST. In our work also, for other values of $x$ (0.5 % by weight and above), the ceramics exhibited mixed phases of only TiO$_2$ and CaTiO$_3$ and the secondary phases reported for the glasses in the system TeO$_2$-CaCu$_3$Ti$_4$O$_{12}$ were not observed [50]. The TeO$_2$ added samples exhibited lower densities than that of the pure CCTO.

Fig.3a shows the SEM picture recorded for the pure CCTO ceramic sintered at 1130$^o$C/2h. These ceramics exhibited fascinating microstructure, in which each grain is surrounded by exfoliate sheets of Cu-rich phase. The EDX studies carried out on the grain and the grain boundary regions clearly indicates that the grain boundary region is rich in Cu. Hence, the microstructural features are broadly described as the stoichiometric grains embedded in a "pool" of Cu-rich phase [26]. It is surprising to see that the microstructure that is evolved for the TeO$_2$ added samples (fig.3(b&c)), this exfoliated sheets of cu-rich phase at the grain boundary regions is very much lesser as compared to that of the pure CCTO. There is a huge difference in the morphology for the pure CCTO and TeO$_2$ added CCTO ceramics. Our results are in agreement with the results obtained for the ZrO$_2$ added CCTO [29], wherein, it has been reported that there is no change in the grain size irrespective of the ZrO$_2$ concentration. Similar results were reported for CaTiO$_3$ addition [30], SrTiO$_3$ doped CCTO ceramics [33], and for the TeO$_2$ doped samples [49], wherein, the grain size does not seems to change so much from sample to sample. In the case of SiO$_2$ addition in CCTO [31], drastic change in the microstructure was observed. From these results, the microstructure evolved for the doped CCTO is dependent on the various parameters like, synthetic route, type and content of doping agent, sintering duration, time and the initial particle size. Hence, the observed microstructure in this work is unique in nature, as the ceramic powders are derived from the complex oxalate precursor route. It is seen from the SEM image (fig.3(b&c)) that the exfoliated sheets of cu-rich phase at the grain boundary regions is very much lesser for the $TeO_2$ added samples as compared to that of the pure CCTO. The EDX analysis also indicated the presence of Ca, Ti, and Te phases in the grains with slight deficiency in Cu atoms in the grain boundaries. These results suggesting that there is a possibility that the CuO segregation formed at the grain boundary regions probably reacted with $TeO_2$. Despite the fact that $TeO_2$ is stable only upto 800°C and it begins to evaporate after that [51,52], it can also get oxidized to tellurates at high temperatures forming Te-Cu-O system [53,54]. For Te-Cu-O system, $CuO_2.TeO_2$ and $CuO.TeO_2$ are the only two stable compounds that could be formed at high temperatures. These compounds are of the nature $CuTe_2O_5$ (orthorhombic)/$CuTeO_3$ (monoclinic) [53,54], for which detailed crystal structures were investigated [55,56]. However, x-ray diffraction pattern recorded for the $TeO_2$ added samples (Fig.2(a-d)) did not show any traces for $CuTeO_5$ (orthorhombic/$CuTeO_3$(monoclinic). This might be due to the fact that, the quantities of $CuTe_2O_5$(orthorhombic/$CuTeO_3$(monoclinic) formed are in proportion to the amount of CuO segregation and the quantities of these individual phases are lower than the detection limits of the XRD technique. Hence, these phases could not be identified by XRD. However, the possibility of formation of orthorhombic $CuTe_2O_5$ /monoclinic $CuTeO_3$ which probably present in the grains cannot be ruled out. The needs to be substantiated with the detailed TEM studies , which is not covered in this work.

The CCTO ceramic sintered at 1130°C/2h has exhibited reasonably high density, these samples were chosen for the dielectric studies. The frequency dependent permittivity ($\varepsilon'_r$) and dielectric loss (D) at room temperature for pure CCTO and $TeO_2$ added samples were shown in the fig. 4(a & b). The permittivity obtained for the $TeO_2$ added samples are lower than that of the pure CCTO. The dielectric content decreased as the $TeO_2$ content increased from 0.5% to 2.0 wt % in the CCTO (fig.4a). The permittivity for the pure CCTO is around 19874 @ 1 KHz whereas, the 2% $TeO_2$ doped samples exhibited a permittivity around 7725 at 1 KHz. The value of permittivity obtained for the 0.5% $TeO_2$ added sample is 8900, which has decreased to 7725 at 1 KHz when the 2% by wt of $TeO_2$ was added to CCTO. Similar trends have been observed by other [29-35] wherein, the addition of doping agents in CCTO, decreased the permittivity and the dielectric loss. It is important to note that, though there is a reduction in the permittivity for the $TeO_2$ added samples in this work, the value obtained is higher than that reported by other [29-35]. The $TeO_2$ added samples exhibited low frequency dispersion in the 100-1MHz range as compared to the pure CCTO samples. The Cu ions in CCTO appear to play a very important role as the segregation of copper oxide at the grain boundaries is believed to be responsible for the high resistance associated with the grain boundary. However, in this work, it is observed that the addition of $TeO_2$ in CCTO forming binary compounds ($CuTe_2O_5$/$CuTeO_3$) as mentioned earlier and reducing the CuO segregation at the grain boundaries and hence reduces the permittivity in CCTO. The SEM analysis very clearly indicated that there is a remarkable difference in the microstructural features for these ceramics and is evidenced that the dielectric response in these ceramics has strong microstructure dependence. The another interesting observations is that, the X-ray powder diffraction obtained for the $TeO_2$ added ceramics showed additional secondary phases of $TiO_2$ and $CaTiO_3$ (CTO) (Fig.2a-d) and grain size smaller than that of the pure CCTO (Fig3a-c). Though the $TiO_2$ rich phase in CCTO reported to exhibit high dielectric response [22], we did not observed any improvement in the dielectric response due to the presence of $TiO_2$ phase. But, as revealed by the XRD (Fig.2a-d), the presence of CTO secondary phases in CCTO is responsible for reducing the permittivity in CCTO. This is in line with the reported literature [34], that the presence of CTO phase in CCTO resulted in

CCTO+CaTiO$_3$ composite, wherein CTO acts as a barrier layer reduces the permittivity. In addition to the above, the absence of CuO segregation at the grain boundaries is also another factor for decreasing the permittivity in the TeO$_2$ added samples. At low frequency, the dielectric loss increases as the TeO$_2$ addition increases (fig.4b), but in the frequency range (10 kHz to 1MHz), all the samples exhibited very low loss value as composed to that of the pure CCTO. The dielectric loss @ 1kHz for the pure CCTO is 0.062 and for the 2% TeO$_2$ added samples, the dielectric loss value is around 0.060. The 2% TeO$_2$ added sample exhibited very low dielectric loss value of 0.037 @ 10 kHz compared to that of pure CCTO. It is also noticed that the dielectric loss value is almost independent of frequency in this range. The dielectric loss did not show any relaxation at low frequency for both pure CCTO as well as for the TeO$_2$ added samples, though there is a relaxation at high frequency observed only for the pure CCTO samples. The doping studies in CCTO were mainly attempted to improve its dielectric loss characteristics by controlling the chemistry and structure of interfacial regions at the grain boundaries [29-34]. In this work, as revealed by the SEM microstructure, the TeO$_2$ addition controlled the CuO segregation at the grain boundaries and thereby reduced the dielectric loss by increasing the grain boundary resistance.

The temperature dependent (50$^o$-150$^o$C) characteristics of permittivity ($\varepsilon_r'$) and dielectric loss (D) were monitored at three different frequencies (1kHz,10kHz,100kHz) for the samples (pure CCTO, 0.5 and 2.0% TeO$_2$ added CCTO) under investigation are depicted in figs. 5 and 6 respectively. The permittivity decreases as the frequency increased from 1kHz to 100kHz in the 50$^o$-150$^o$C temperature range. The permittivity increases steadily with increasing temperature for all the samples. The permittivity value at 1kHz has increased from 23428 to 35964 as the temperature increased from 50$^o$C to 150$^o$C for the pure CCTO sample (fig.5a) and for the 0.5% TeO$_2$ added sample (fig.5b), the increase is from 8900 to 14550. In the case of 2.0% $TeO_2$ added sample (fig.5c), it has increased from 7527 to 14537 when the temperature is increased from $50^oC$ to $150^oC$. The temperature dependent ($50^o$-$150^oC$) characteristics of dielectric loss (D) at various frequencies for all the samples (pure CCTO, 0.5 and 2.0% $TeO_2$ added) are shown in the fig.6. In the case of pure CCTO (fig.6a), the frequency independent dielectric loss was observed upto $75^oC$ and subsequently it decreased with increasing frequency. The dielectric loss increases rapidly beyond $100^oC$. The $TeO_2$ added samples also exhibited similar behaviour. The dielectric loss for the pure CCTO is increased from 0.099 to 0.80 @ 1kHz, similarly, dielectric loss has increased from 0.043 to 0.92 for the 2.0% $TeO_2$ added sample (fig.6c), indicating the dominance of leakage current at elevated temperatures.

In order to have an insight into the relatively low dielectric loss observed for the $TeO_2$ added samples, the ac conductivity measurements were carried out in the frequency of interest for all the three samples under investigation. The frequency dependence of the ac conductivity (in log-log scale) at room temperature is shown in fig 7. The conductivity for all the three samples is found to increase with increasing frequency. Over all, it has been observed that the $TeO_2$ added CCTO has exhibited lower conductivity behaviour compared to that of the CCTO. This is a clear indication that the $TeO_2$ addition increases the resistance at grain boundary by reaction with the CuO segregation formed at the grain boundaries, which was visible in the SEM micrographs recorded. These data indicate that $TeO_2$ added CCTO maintains lower dielectric loss by increasing the grain boundary resistance.

**Acknowledgement**

The management of Central Power Research Institute is acknowledged for the financial support (CPRI Project No. R-DMD-01/1415).

*Conclusions*

Giant dielectric CCTO ceramic with low dielectric loss has been developed by the addition of $TeO_2$ into the nano ceramics of CCTO, derived from the complex oxalate precursor route. The addition of $TeO_2$ in CCTO has reduced the CuO segregation at the grain boundary and hence, there is a remarkable difference in the microstructural features by the addition of $TeO_2$. The ceramics thus prepared has exhibited permittivity as high as 7387 @ 10 kHz and low dielectric loss value of 0.037 @ 10 kHz, which can be exploited for the high frequency capacitors application.

## FIGURE CAPTIONS

Figure. 1. X-ray diffraction patterns of (a) as prepared $CaCu_3Ti_4O_{12}$ (CCTO) nanopowders, CCTO pellets sintered (b) at $1100^oC/2h$, (c) at $1130^oC$ exhibiting single phase nature and (d) ICDD data file card no. 01-075-1149 for CCTO.

Figure. 2. X-ray diffraction patterns of pellets sintered at $1130^oC$, (a) 0.5 wt % $TeO_2$ in CCTO, (b) 1.0 wt % $TeO_2$ in CCTO, (c) 1.5 wt % $TeO_2$ in CCTO and (d) 2.0 wt % $TeO_2$ in CCTO.

Figure. 3. Scanning electron micrographs of the CCTO sintered at $1130^oC/4h$, (a) showing the exfoliated sheets of Cu rich phase at the grain boundary region (b) 0.5 wt % $TeO_2$ in CCTO and (C) 2.0 wt % $TeO_2$ in CCTO.

Figure 4. Frequency dependence of room-temperature (a) permittivity and (b) dielectric loss for the pellets sintered at $1130^oC/4h$ as a function of $TeO_2$ content.

Figure. 5 Temperature dependence of permittivity measured at selected frequencies for (a) CCTO (b) 0.5 % $TeO_2$ in CCTO and (c) 2.0 % $TeO_2$ in CCTO.

Figure.6. Temperature dependence of dielectric loss (D) measured at selected frequencies for (a) CCTO (b) 0.5 % $TeO_2$ in CCTO and (c) 2.0 % $TeO_2$ in CCTO.

Figure.7. Frequency dependence of ac conductivity measure at 300K for CCTO, and the $TeO_2$ added CCTO ceramic samples.

.

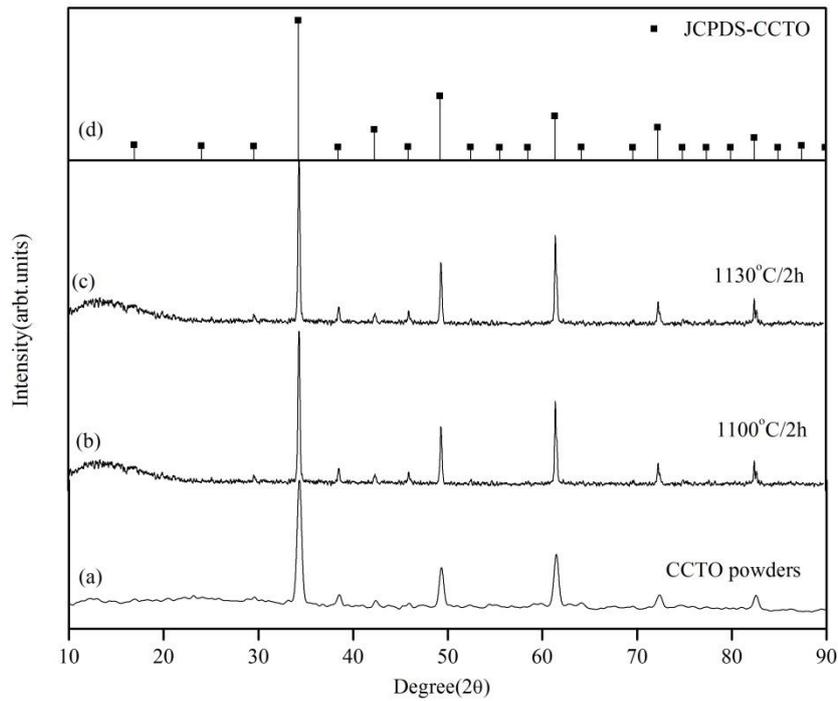

Figure. 1. X-ray diffraction patterns of (a) as prepared $CaCu_3Ti_4O_{12}$ (CCTO) nanopowders, CCTO pellets sintered (b) at 1100°C/2h , (c) at 1130°C exhibiting single phase nature and (d) ICDD data file card no. 01-075-1149 for CCTO.

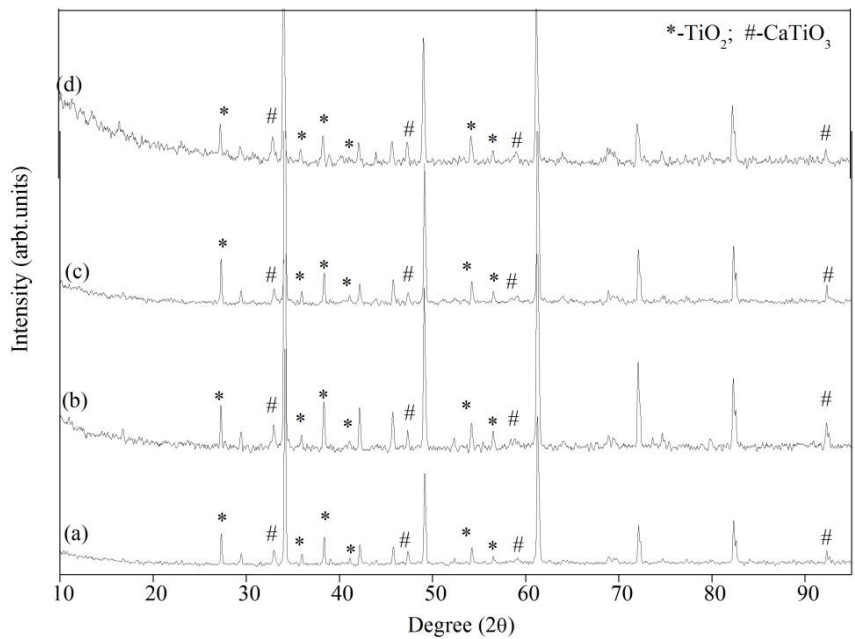

Figure. 2. X-ray diffraction patterns of pellets sintered at 1130°C, (a) 0.5 wt % $TeO_2$ in CCTO, (b) 1.0 wt % $TeO_2$ in CCTO , (c) 1.5 wt % $TeO_2$ in CCTO and (d) 2.0 wt % $TeO_2$ in CCTO.

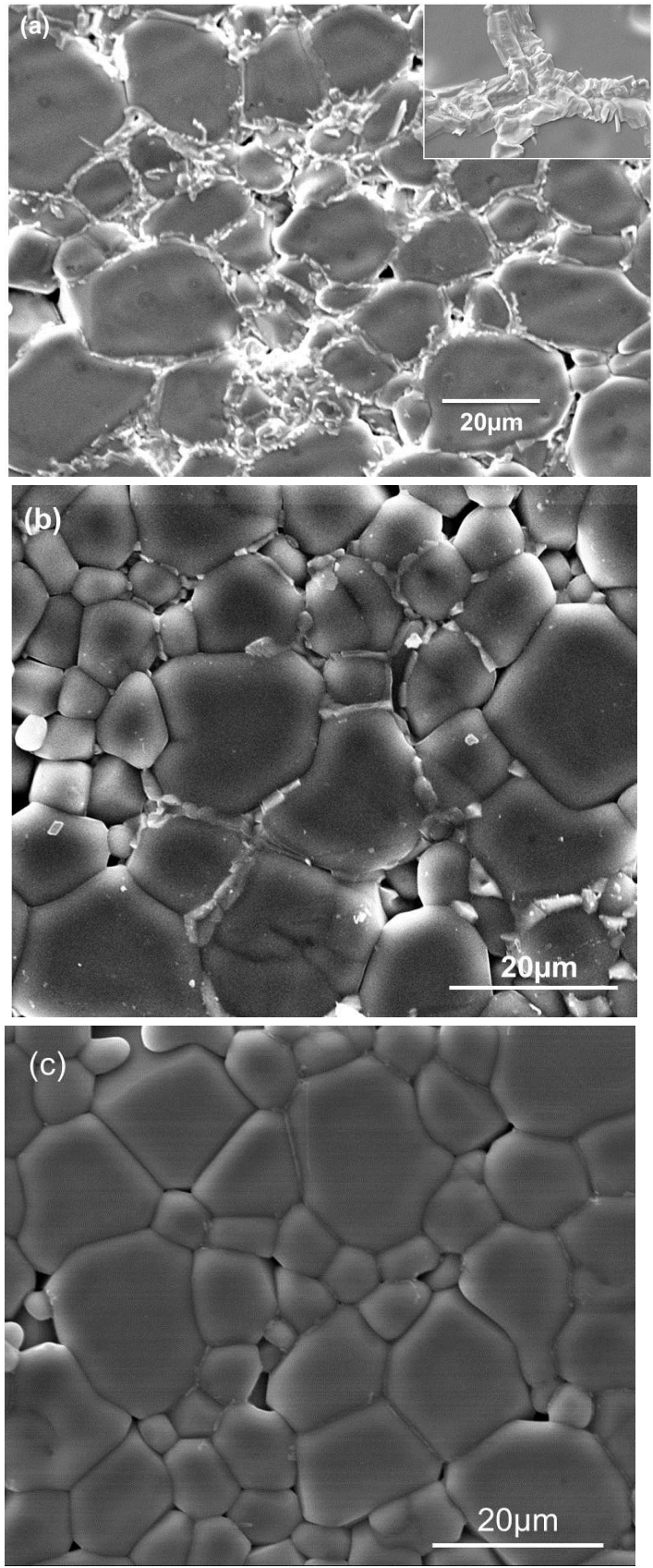

Figure. 3. Scanning electron micrographs of the CCTO sintered at 1130°C/4h, (a) showing the exfoliated sheets of Cu rich phase at the grain boundary region (b) 0.5 wt %TeO$_2$ in CCTO and (C) 2.0 wt % TeO$_2$ in CCTO.

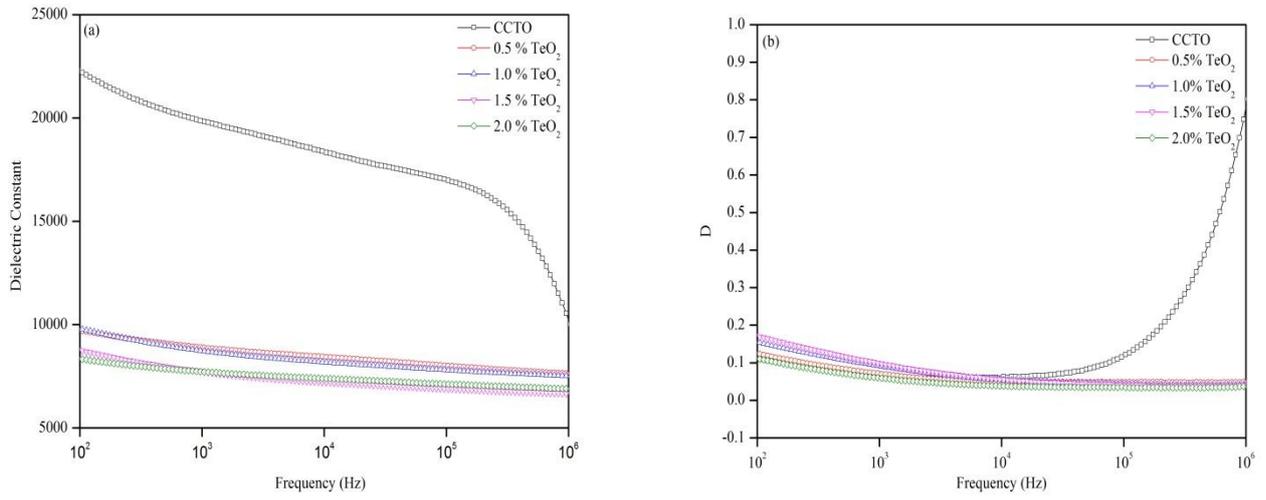

Figure 4. Frequency dependence of room-temperature (a) permittivity and (b) dielectric loss for the pellets sintered at 1130°C/4h as a function of $TeO_2$ content.

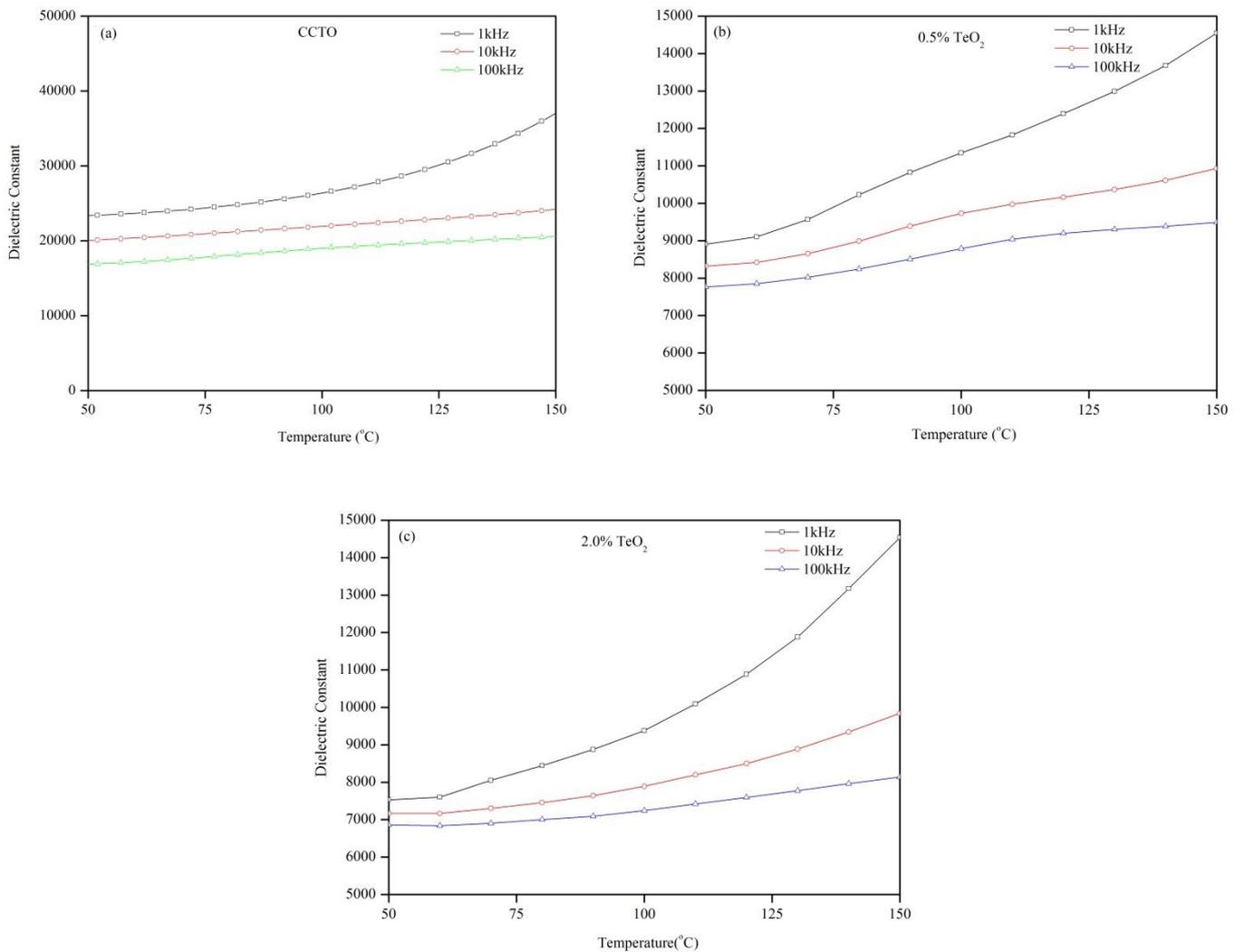

Figure. 5 Temperature dependence of permittivity measured at selected frequencies for (a) CCTO (b) 0.5 % $TeO_2$ in CCTO and (c) 2.0 % $TeO_2$ in CCTO.

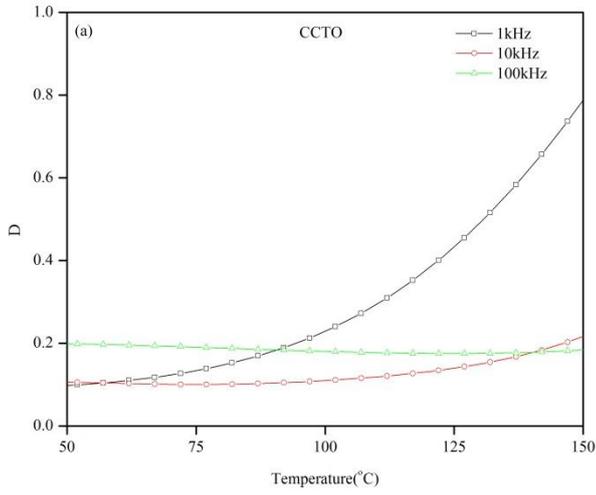
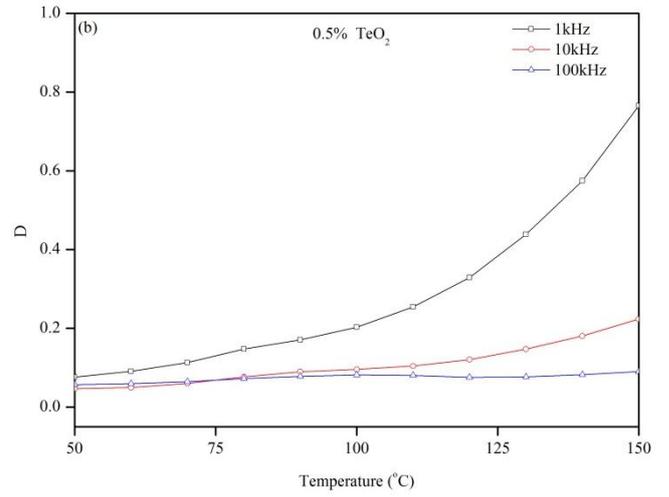
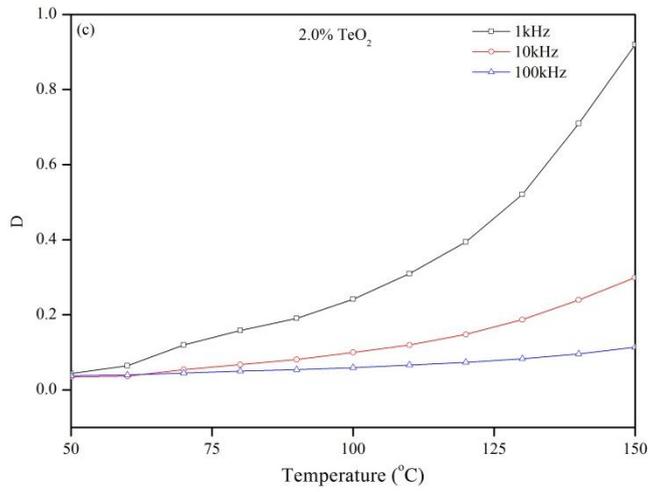

Figure.6. Temperature dependence of dielectric loss (D) measured at selected frequencies for (a) CCTO (b) 0.5 % $TeO_2$ in CCTO and (c) 2.0 % $TeO_2$ in CCTO.

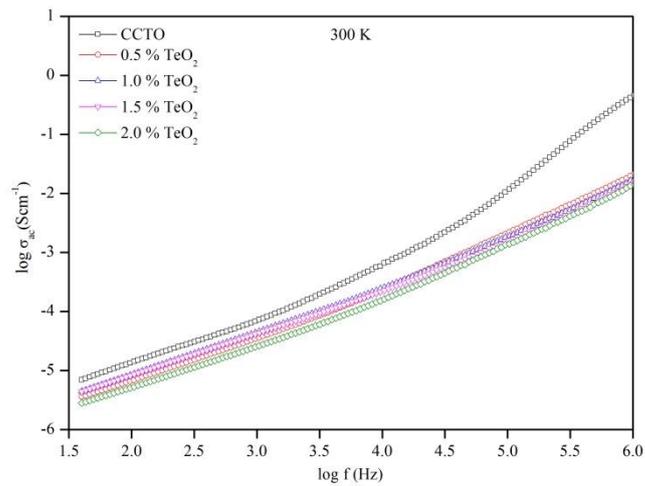

Figure.7. Frequency dependence of ac conductivity measure at 300K for CCTO, and the $TeO_2$ added CCTO ceramic samples.